\documentclass[preprint,aps,floats,showpacs,amssymb,tightenlines]{revtex4}
\usepackage{epsfig}

\usepackage{amsfonts}
\usepackage{amsmath}
\newcommand{\beq}{\begin{equation}}
\newcommand{\eeq}{\end{equation}}
\newcommand{\be}{\begin{equation}}
\newcommand{\ee}{\end{equation}}
\newcommand{\beqs}{\begin{eqnarray}}
\newcommand{\eeqs}{\end{eqnarray}}
\newcommand{\bea}{\begin{eqnarray}}
\newcommand{\eea}{\end{eqnarray}}

\newcommand{\bml}{\begin{mathletters}}
\newcommand{\eml}{\end{mathletters}}

\newcommand{\beast}{\begin{eqnarray*}}
\newcommand{\eeast}{\end{eqnarray*}}

\newcommand{\vphi}{\varphi}

\begin{document}

%\tighten
%\draft

%%%%%%%%%%%%%%%%%%%%%%%%%%%%%%%%%%%%%%%%%%%%%%%%%%%%%%%%%%%%%%%%%%%%%%%%%%

%\rightline{Preliminary Draft}

\title{Charged-rotating black holes and black strings
in higher dimensional
Einstein-Maxwell theory with a positive
cosmological constant}
\renewcommand{\thefootnote}{\fnsymbol{footnote}}

\author{Y. Brihaye \footnote{Brihaye@umh.ac.be} , T. Delsate}
\affiliation{Dep de Math\'ematiques et Physique Th\'eorique,
Universit\'e de Mons, Place du Parc, 7900 Mons, Belgique}

\date{\today}
\setlength{\footnotesep}{0.5\footnotesep}

%%%%%%%%%%%%%%%%%%%%%%%%%%%%%%%%%%%%%%%%

\begin{abstract}We present arguments for the existence of charged,
rotating black holes in $d=2N+1$ dimensions, with $d\geq 5$
with a positive cosmological constant.
These solutions posses both,
a regular horizon and a cosmological horizon of spherical topology and
have $N$ equal-magnitude angular momenta. They
approach asymptotically the de Sitter spacetime background.
The counterpart equations for  $d=2N+2$ are investigated, by assuming
that the fields are independant of the extra dimension $y$, leading
to black strings solutions.  These
solutions are regular at the event horizon.  The asymptotic form of the metric 
is not the de Sitter form and exhibit a naked 
singularity at finite proper distance.
\end{abstract}

\maketitle
%%%%%%%%%%%%%%%%%%%%%%%%%%%%%%%%%%%%%%%%%%%%%%%%%%%%%%%%%%%%%%%%%%%%%%%%%%%%%%%%%%%%%%%%%%%%%%%%%%
%%%%%%%%%%%%%%%%%%%%%%%%%%%%%ICI%%%%%%%%%%%%%%%%%%%%%%%%%%%%%%%%%%%%%%%%%%%%%%%%%%%%%%%%%%%%%%%%%%%
\section{Introduction}
Recently, there was a lot of interest in black holes and black strings solutions
is space-time with arbitrary dimensions and with a cosmological constant $\Lambda$.
In the case of a negative cosmological constant, the interest of these
solutions is related to the correspondance between gravitating fields in an AdS space-time and 
the conformal field theory on the boundary of the AdS space-time \cite{Witten:1998qj,Maldacena:1997re}.

Black-string solutions denote a string like generalisation of 4-D black hole solution of d-dimensional Einstein gravity characterized 
with an event horizon of topology  $S_{d-3} \times S_1$ \cite{bs}.
Black string for $d = 5$ and $\Lambda < 0$ they were first considered in \cite{Copsey:2006br}
and  then generalized to $d$ dimensions   in \cite{mrs}.  
The charged counterparts of these solutions for the minimal Einstein-Maxwell (EM) 
gravity have been obtained recently \cite{brs}. For space-times of even dimensions $d\geq 6$, the $S_{d-3}$ part
of the metric can be deformed by using the ideas of \cite{Gibbons:2004js}  in such a way that some rotating black string
 solutions can be constructed while using normal differential equations only.
Rotating counterparts of the black strings of \cite{mrs}, constructed along these lines, 
are also obtained in \cite{brs}. Another direction of investigation of these objects
is non-uniform black strings, 
where a non trivial dependence on the extra-dimension is required \cite{wiseman}. Recently, $\Lambda=0$
rotating nonuniform solutions have been considered as well \cite{kkr}.

Generalizing the solutions of Tangerlini \cite{tangherlini} and of Myers-Perry \cite{mp},
higher dimensional charged, rotating black holes were constructed in \cite{Kunz:2006eh}
with asymptotically flat space-time using the symmetries of \cite{Gibbons:2004js}.
Very recently, rotating black holes in Einstein-Maxwell (EM) theory were constructed in \cite{knlr} for 
odd space-time dimensions $d \geq 5$ and $\Lambda < 0$. 

The litterature investigating higher dimensional  black holes and black strings 
with a positive cosmological constant is by far less abundant, although the problem
desserves to be investigated for several reasons. Namely~: 
\begin{description}
\item{i)} the recent experiments are rather consistent
with a cosmological constant of the positive sign, 
\item{ii)} as observed e.g. in \cite{bhr}, a positive cosmological constant can have important
consequences on the physical properties of some classical solutions, these could play a role in inflation
\item{iii)} it is mathematically interesting to see whether 
 the  solutions available for $\Lambda < 0$ can be analytically continued for $\Lambda >0$,
\item{iv)} it is numerically  challenging
since we expect a cosmological horizon to occur at an intermediate value of the radial variable of space-time,
\item{v)} connections between quantum gravity in DeSitter space-time and conformal field theory on the 
boundary exist, see e.g. \cite{kv},
\item{vi)} finally,
such solutions would extend the pattern of already known solutions of Einstein equations in higher
dimensional space-times.
\end{description}
 Up to now, 
 several charged and/or rotating black holes solutions of the Einstein equations coupled to an electromagnetic
 fields  in space-times  with $d\geq 4$ (and $\Lambda > 0$) are known \cite{ks}, \cite{clp};
  they are constructed with a Chern-Simons term which appears naturally in the bosonic sector of
  minimal five-dimensional supergravity. 
  Vacuum solutions of the vacuum 5-dimensional Einstein gravity with $\Lambda > 0$ were obtained in \cite{bd}, 
  but a systematic study of (rotating) solutions of the minimal Einstein-Maxwell
  theory (with $\Lambda > 0$) has, to our knowledge, not yet been adressed. Let us mention that static
  black rings  solutions have also been constructed for 5-d DeSitter supergravity theory \cite{chinese}.
  Solutions of the Einstein-Maxwell-dilaton system were are considered in \cite{charmousis} with both
  signs of the cosmological constant.

Considering the minimal theories of Einstein  and Einstein-Maxwell gravity
 with the same kinds of symmetries of the fields as in \cite{knlr} and \cite{brs} but with a positive
cosmological constant can lead to solutions presenting drastic
differences with respect to the case of negative cosmological constant, 
although the equations to solve are basically similar. 
In this paper, we will reconsider the equations of \cite{knlr,brs} for positive values
of the cosmological constant and construct several families of solutions of these equations. The
paper is organized as follow~:
Sect. 2 is devoted to rotating, charged black holes in odd dimensions. The ansatz, the equations,
 the boundary conditions and the numerical results are  presented successively. Non-rotating and
rotating black strings (in even dimensions) are described in Sect. 3, following a similar pattern.
The results are summarized in Sect. 4.

\section{Black holes}

We consider the Einstein-Maxwell action with a positive cosmological 
constant $\Lambda$
\begin{eqnarray}
\label{action-grav}
I=\frac{1}{16 \pi G}\int_M~d^dx \sqrt{-g} (R - 2\Lambda-F_{\mu \nu}F^{\mu \nu})
-\frac{1}{8\pi G}\int_{\partial M} d^{d-1}x\sqrt{-h}K,
\end{eqnarray}
in a $d-$dimensional spacetime.
The last term in  (\ref{action-grav}) is the Hawking-Gibbons surface
term \cite{Gibbons:1976ue},
which
is required in order to have a well-defined variational principle. The factor 
$K$ represents the trace
of the extrinsic curvature for the boundary $\partial\mathcal{M}$ and
$h$ is the induced
metric of the boundary.
Along with many authors, we note  $\Lambda= \pm (d-2)(d-1)/(2\ell^2)$, 
the sign $+$ will be considered throughout this paper.
%%%%%(\textbf{Terence : + sign for} $\Lambda>0$ \textbf{and - sign for} $\Lambda>0$).

%%%%%%%%%%%%%%%%%%%%%%%%%%%%%%%%%%%%%%%%%%%%%%%%%%%%%%%%%%%%%%%%%%%%
\subsection{The ansatz}
%%%%%%%%%%%%%%%%%%%%%%%%%%%%%%%%%%%%%%%%%%%%%%%%%%%%%%%%%%%%%%%%%%%%

To obtain rotating black hole solutions,
representing charged $U(1)$ generalizations of the vacuum
 solutions discussed in \cite{Gibbons:2004js},
we consider space-times with odd dimensions and possessing $N=(d-1)/2$ commuting Killing vectors
$\eta_k=\partial_{\varphi_k}$.

We use a parametrization for the metric,
corresponding to a generalization of the Ansatz
previously used for asymptotically flat solutions \cite{Kunz:2006eh}
\begin{eqnarray}
 ds^2 = -b(r)dt^2 +  \frac{ dr^2}{f(r)} +
g(r)\sum_{i=1}^{N-1}
  \left(\prod_{j=0}^{i-1} \cos^2\theta_j \right) d\theta_i^2
  \nonumber \\
 + h(r) \sum_{k=1}^N \left( \prod_{l=0}^{k-1} \cos^2 \theta_l
  \right) \sin^2\theta_k \left( d\vphi_k - w(r)
  dt\right)^2
\nonumber
 \\
\label{metric}
 +p(r) \left\{ \sum_{k=1}^N \left( \prod_{l=0}^{k-1} \cos^2
  \theta_l \right) \sin^2\theta_k  d\vphi_k^2 \right.
  -\left. \left[\sum_{k=1}^N \left( \prod_{l=0}^{k-1} \cos^2
  \theta_l \right) \sin^2\theta_k   d\vphi_k\right]^2 \right\} \ ,
  \label{metric}
\end{eqnarray}
In the above formula  $\theta_0 \equiv 0$ and  $\theta_N \equiv \pi/2$ are assumed;
the  non trivial angles have      $\theta_i \in [0,\pi/2]$
for $i=1,\dots , N-1$, while
 $\vphi_k \in [0,2\pi]$ for $k=1,\dots , N$. The functions $b,f,h,g,w$ depend on the variable $r$.
The consistency of the ansatz imposes $p(r)=g(r)-h(r)$.

The parametrization of the U(1) potential, consistent with the symmetries
of the  line element (\ref{metric}) is
\begin{eqnarray}
\label{Maxwell}
 A_\mu dx^\mu=V(r)dt+a_\varphi(r) 
\sum_{k=1}^N
\left( \prod_{l=0}^{k-1} \cos^2\theta_l \right) 
\sin^2\theta_k d\vphi_k
\end{eqnarray}
Here, the electric and magnetic potentials $V(r)$ and  $a_\varphi(r)$ also depend on $r$.
The Einstein-Maxwell equations lead to a consistent system of ordinary differential equations
in the different radial functions.

Without fixing a metric gauge, the ansatz presented above
leads to following  reduced  
Lagrangean of the EM system, say  $L=L_g-L_M$. 
The Einstein part $L_g$ and Mawxell part $L_M$ read respectively~:
\begin{eqnarray}
\label{Lg}
L_g=(d-3)g^{\frac{(d-7)}{2}}\sqrt{\frac{bh}{f}}((d-1)g-h)
+\frac{1}{2}\sqrt{fhb}g^{\frac{(d-3)}{2}}
(\frac{b'}{b}+(d-3)\frac{g'}{g})(\frac{h'}{h}+(d-3)\frac{g'}{g})
\\
\nonumber
-\frac{1}{4}(d-2)(d-3)\sqrt{fhb}g^{\frac{(d-7)}{2}}g'^2
+\frac{1}{2}g^{\frac{(d-3)}{2}}h\sqrt{\frac{fh}{b}}w'^2
-\frac{(d-2)(d-1)}{\ell^2}g^{\frac{(d-3)}{2}}\sqrt{\frac{bh}{f}}
\end{eqnarray}
\begin{eqnarray}
\label{LM}
L_M=\frac{g^{\frac{(d-7)}{2}}}{\sqrt{bfh}}
\bigg(
2b\left(
2(d-3)a_\varphi^2h+fg^2a_\varphi'^2
\right)
-2fg^2h(w a_\varphi'+V')^2
\bigg).
\end{eqnarray}
%

%%%%%%%%%%%%%%%%%%%%%%%%%%%%%%%%%%%%%%%%%%%%%%%%%%%%%%%%%%%%%%%%%%%%
\subsection{The equations}
%%%%%%%%%%%%%%%%%%%%%%%%%%%%%%%%%%%%%%%%%%%%%%%%%%%%%%%%%%%%%%%%%%%%
The EM equations obtained from the ansatz discussed above can be obtained in a standard way.
For the numerical construction of the solutions, the "`metric gauge"' has to be fixed;  we find it is convenient to fix  
it by imposing  $g(r)=r^2$.
With this gauge, the following field equations are found after an algebra~:
\begin{eqnarray}
\label{ec2}
f'
+\frac{f}{(d-2)}
\bigg(-\frac{rh}{2b}w'^2
+\frac{2r}{ b}V'^2
+\frac{4rw}{b}a_\varphi 'V'
+\frac{h'}{h}(1-\frac{rb'}{2b})
-2r(\frac{1}{h}-\frac{w^2}{b})a_\varphi'^2
+\frac{b'}{b}
\\
\nonumber
+\frac{(d-1)(d-4)}{r}
\bigg)
+\frac{1}{(d-2)r^3}
((3d-5)h+4(d+1)a_\varphi^2-(d-1)^2r^2)+\frac{(d-1)r}{\ell^2}
=0,
\end{eqnarray}
\begin{eqnarray}
\label{ec1}
b''+
 \frac{1}{d-2}
\bigg(
 4(5-2d)w a_\varphi'V'
 +\frac{(d-3)}{2h}b'h'
 -\frac{2(d-3)b}{2rh}h'
 -2(2d-5)V'^2
\\
\nonumber
 +\frac{1}{2}(3-2d)hw'^2
-2(\frac{b}{h}+(2d-5)w^2)a_\varphi'^2
+(d-2)\left(\frac{b'f'}{2f}-\frac{b'^2}{2b}\right)
+\frac{(d-3)^2}{r}b'
\\
\nonumber
-\frac{(d-3)b}{r^4f}(12 a_\varphi^2+h)
+\frac{(d-3)b}{r^2}\left(\frac{d-1}{f}+4-d\right)
+\frac{(d-1)(d-2)b}{\ell^2f}
\bigg)=0,
\end{eqnarray}
\begin{eqnarray}
\label{ec3}
h''
+\frac{1}{(d-2)}
\bigg(
\frac{(2d-5)h^2}{2b}w'^2
+\frac{2h}{b}V'^2
+\frac{4hw }{b}a_\varphi'V'
-\frac{(d-2)h'}{2}(\frac{h'}{h}-\frac{f'}{f})
\\
\nonumber
+\frac{(d-3)}{2b}b'h'
+\frac{(d-3)^2}{r}h'
+2(\frac{hw^2}{b}+2d-5)a_\varphi'^2
-\frac{(d-3)h}{rb}b'
-\frac{(d-3)(2d-3)h^2}{r^4f}
\\
\nonumber
-\frac{12(d-3)a_\varphi^2h}{r^4f}
-\frac{(d-3)(d-4)h}{r^2}
+\frac{(d-1)h}{f}(\frac{d-2}{\ell^2}-\frac{d-3}{r^2})
\bigg)=0,
\end{eqnarray}
\begin{eqnarray}
\label{ec4}
w''
-\frac{4w }{h}a_\varphi'^2
-\frac{4a_\varphi'V'}{h}
+\frac{(d-3)w'}{r}
+\frac{1}{2}\left(-\frac{b'}{b}+\frac{f'}{f}+\frac{3h'}{h}\right)w'=0,
\end{eqnarray}
%
%
%
%
%%%%%%%%%%%\begin{eqnarray}
%%%%%%%%%%\label{ec5}
%%%%%%%%%%\frac{(d-3)rh}{b}b'
%%%%%%%%%%-2(1-\frac{hw^2}{b})r^2a_\varphi'^2
%%%%%%%%%%%+(d-3)rh'
%%%%%%%%%%%%+\frac{r^2b'h'}{2b}
%%%%%%%%%%%+\frac{4r^2hw}{b}a_\varphi'V'
%%%%%%%%%%%+\frac{2r^2h}{b}V'^2
%%%%%%%%%%%%%%+\frac{r^2h^2}{2b}w'^2
%%%%%%%%%%%%%%%\\
%%%%%%%%%%%%%%\nonumber
%%%%%%%%%%%%%+(d-3)(d-4)h
%%%%%%%%%%%%%%-\frac{(d-1)h}{f}(d-3+\frac{(d-2)r^2}{\ell^2})
%%%%%%%%%%%%%%+\frac{(d-3)h(4a_\varphi^2+h)}{r^2f}=0,
%%%%%%%%%%%%%%%\end{eqnarray}
%
%
for the gravity part, and
%%\footnote{Note the existence of a one extra-equation (a constraint), 
%%which is satisfied
%%automatically as a consequence of the consistency of the ansatz.}, and
%
%
\begin{eqnarray}
\label{ec6}
V''
-\frac{w}{b}b'a_\varphi'
+\frac{w}{h}a_\varphi'h'
+\frac{1}{2}(\frac{2(d-3)}{r}-\frac{b'}{b}+\frac{f'}{f}+\frac{h'}{h})V' \nonumber \\
+(1+\frac{hw^2}{b})a_\varphi'w'
+\frac{hw}{b}V'w'
+\frac{2(d-3)a_\varphi hw}{r^4f}=0,
\end{eqnarray}
\begin{eqnarray}
\label{ec7}
a_\varphi''
+\frac{1}{2}(\frac{2(d-3)}{r}+\frac{b'}{b}+\frac{f'}{f}-\frac{h'}{h})a_\varphi'
-\frac{h}{b}(wa_\varphi'+V')w'
-\frac{2(d-3)a_\varphi h}{r^4f}=0.
\end{eqnarray}
for the U(1) potentials.  
It can easily be seen that the  equations of motion  present the
first integral
\begin{eqnarray}
\label{fiwbh}
 g^{\frac{(d-3)}{2} }\sqrt{\frac{fh}{b}}(wa_\varphi'+V')=(d-3)q.
\end{eqnarray} 
Thus, similar to the asymptotically flat case  \cite{Kunz:2006eh} case, the electric potential 
can be eliminated from the equations (\ref{ec6}) 
by making use of the first integral
(\ref{fiwbh}). The cosmological constant parameter can be arbitrarily rescaled by a
rescaling of the radial variable $r$ and of the fields $h$ and $w$. In this section,
we  use this arbitrariness to choose $r_h=1$ without loosing generality.

{\bf Known solutions}
The vacuum black holes discussed in \cite{Gibbons:2004js}
are recovered for a vanishing gauge field and
\begin{eqnarray}
\label{vacuum}
 f(r)=1
-\frac{r^2}{\ell^2}
-\frac{2M\Xi}{r^{d-3}}
+\frac{2Ma^2}{r^{d-1}},~ 
h(r)=r^2(1+\frac{2Ma^2}{r^{d-1}}),~ \nonumber  \\
w(r)=\frac{2Ma}{r^{d-3}h(r)},~~
g(r)=r^2,~~ b(r)=\frac{r^2f(r)}{h(r)},
\end{eqnarray}
where $M$ and $a$ are two constants related to the solution's mass and 
angular momentum and $\Xi=1+a^2/\ell^2$.
\subsection{Constraint of regularity about the horizon}
We are interested in black hole solutions, with an horizon
located at $r=r_h$.
The solutions can be expanded in the neighbourhood of the horizon in the same 
was as in the case of a negative cosmological constant \cite{knlr}, i.e.
\begin{eqnarray}
b(r)=b_1(r-r_h)+O(r-r_h)^2,
~~f(r)=f_1(r-r_h)+O(r-r_h)^2, \nonumber
\\
h(r)=h_0+h_1(r-r_h)+O(r-r_h)^2,
w(r)=w_h+w_1(r-r_h)+O(r-r_h)^2, \nonumber \\
\nonumber a_\varphi(r)=a_0+a_1(r-r_h)+O(r-r_h)^2,~~
V(r)=V_0+V_1(r-r_h) + O(r-r_h)^2
\end{eqnarray}
%%with 
%%\begin{eqnarray}
%%&f_1=\frac{b_1(d-2)(d-1)r_h^4+b_1(-4a_0^2(d-1)+(d-2)(-2h_0+(d-1)r_h^2))\ell^2}
%%{r_h^3\ell^2(b_1(d-2)+2r_h(v_1+a_1w_h)^2)},
%%\\
%%\nonumber
%%&a_1=\frac{((d-3) h_0r_h^{-d-4}}{\sqrt{b_1f_1^3h_0}}(2a_0\sqrt{b_1f_1h_0}r_h^d+f_1qr_h^7w_1)~,
%%~
%%v_1=\frac{(d-3)h_0r_h^{-d-4}}{\sqrt{b_1f_1^3h_0}}
%%(-2a_0\sqrt{b_1f_1h_0^3}r_h^d w_h+f_1 q r_h^7(b_1-h_0w_1w_h)).
%%\end{eqnarray}

Although it is not clear if the thermodynamical properties are well defined with the presence of the cosmlogical horizon, one may still define them in the standard way. The Hawking 
temperature and the event horizon area of these solutions can be obtained in a standard way, leading to 
\begin{eqnarray}
T_H=\frac{\sqrt{ f_1b_1}}{4\pi},~~
A_H=V_{d-2}r_h^{d-2}.
\end{eqnarray}
Along with \cite{knlr}, we also write the mass and the angular velocity at the horizon defined by
means of the appropriate Komar integrals~:
\be
\label{komar}
   M_H = \frac{V_{d-2}}{8 \pi G_d} \sqrt{\frac{f h g^2}{b}} (b' - h w w')\vert_{r=r_h} \ \ , \ \ 
   J_H = \frac{V_{d-2}}{8 \pi G_d} 2\sqrt{\frac{f g^2 h^3}{b}} \  w' \vert_{r=r_h}
\ee
where $V_{d-2}$ denotes the area of the $d-2$ dimensional sphere.
These quantities can be easily evaluated from the numerical solutions.

For the solutions to be regular at the horizon $r_h$ (or at the cosmological horizon $r_c$), 
the equation for $h$ leads to the  condition $\Gamma_1(x=r_h) = 0$ with
\beqs
  \Gamma_1(x) \equiv && 8 b' h^2 (12 a^2 + 7 h) + 4 x b' h h' (12 a^2 + 5 h) \nonumber \\
                     &-& 32 b' h^2 x^2 + 8 x^3 b' h (f' h - 4 h')         \nonumber \\
                     &+& 2 h x^4 ( \frac{24}{\ell^2} b' h^2 - f' 
                     (4 (a')^2 h w^2 + 8 a' h w V' + b' h' + 5 h^2 (w')^2 + 4 h (V')^2  )   
                     )                     \nonumber \\
                     &+&  h' x^5 ( - \frac{24}{\ell^2} b' h^2 - f' 
                     (4 (a')^2 h w^2 + 8 a' h w V' - b' h' -  h^2 (w')^2 + 4 h (V')^2  )   
                     )   
\eeqs
where we posed $a_{\phi} \equiv a$.
The value $f'(x_h)$ can be extracted from the equation for $f$, giving
\beq
      f'(x_h) = \frac{4 b_1 h_0}{x_h^3} 
      \frac{6 x_h^4/ \ell^2 + 8 x_h^2 - 12 a_0^2 - 5 h_0}
      {8 b_1 h_0 + x_h(4h_0(V_1+a_1 w_0)^2 - b_1 h_1 - h_0^2 w_1^2)}
\eeq
In the same way,  the two Maxwell lead to a single condition $\Gamma_2(x = r_h) = 0$ with
\beq
\Gamma_2 (x) \equiv 4 a b' h + x^4 f'(h w' V' + a' h w w' - a' b')
\eeq

%
%%%%%%%%%%%%%%%%%%%%%%%%%%%%%%%%%%%%%%%%%%%%%%%%%%%%%%%%%%%%%%%%%%%%
\subsection{The asymptotics and global charges}
%%%%%%%%%%%%%%%%%%%%%%%%%%%%%%%%%%%%%%%%%%%%%%%%%%%%%%%%%%%%%%%%%%%%

In this section, we follow the lines of \cite{knlr} to present the global charges
characterizing the solutions asymptotically. This uses a formalism developped namely in 
\cite{Balasubramanian:1999re,Brown:1993} and also used in \cite{bhr,bhrs} 
The metric functions have the following asymptotic behaviour
in terms of three arbitrary constants $\alpha,~\beta$ and $\hat J$
\begin{eqnarray}
\label{asym}
b(r)=- \frac{r^2}{\ell^2}+1+ \frac{\alpha}{r^{d-3}} +O(1/r^{2d-6}),
~~~
f(r)=- \frac{r^2}{\ell^2}+1+\frac{\beta}{r^{d-3}} +O(1/r^{d-1}),
\\
\nonumber
h(r)=  r^2(1+  \frac{\beta-\alpha}{r^{d-1}} +O(1/r^{2d-4})),
~~~
w(r)=    \frac{\hat J}{r^{d-1}} +O(1/r^{2d-4 }),
\end{eqnarray}
The asymptotic expression of the gauge potential is similar to the asymptotically flat case
\begin{eqnarray}
 V(r)=- \frac{q}{r^{d-3}} +O(1/r^{2d-4 }),~~
 a_\varphi(r)= \frac{\hat \mu}{r^{d-3}} +O(1/r^{2d-4}) .
\end{eqnarray}
 The mass-energy of the solutions and angular momentum  associated with an angular direction is
\begin{eqnarray}
\label{grav-charges}
 E=\frac{V_{d-2}}{16\pi G_d}(\beta-(d-1) \alpha),
~~J=\frac{V_{d-2}}{8\pi G_d}\hat J \ .
\end{eqnarray}
%%%
The above relations can be proven by using a background subtraction approach
or the counterterm formalism \cite{Balasubramanian:1999re,Brown:1993,bhrs}.

The electric charge and the magnetic moment of the solutions are given by
\begin{eqnarray}
\label{gauge-charges}
 Q=\frac{(d-3)V_{d-2} }{4\pi G_d}q,~~~\mu=\frac{(d-3)V_{d-2}}{4\pi G_d}\hat \mu~.
\end{eqnarray}

The ansatz in $f,~b,~h$ has to advantage to present a 
direct connection with the closed form vacuum rotating solution.
%%%%%%%%%%%%%%%%%%%%%%%%%%%%%%%%%%%%%%%%%%%%%%%%%%%%%%%%%%%%%%%%%%%%
%%%\subsection{Expansion about the horizon}
%%%%%%%%%%%%%%%%%%%%%%%%%%%%%%%%%%%%%%%%%%%%%%%%%%%%%%%%%%%%%%%%%%%%

\subsection{Numerical results}
The system of equations above does not admit, to our knowledge, explicit solutions for
generic values of $w_h$ and $V_h$. We therefore relied on a numerical method to
construct solutions. We solved the equations in the case $d=5$ and we hope that this case
catches the qualitative properties of the pattern of the solutions; the numerical solver
Colsys \cite{colsys} was used throughout this paper. 

The positive cosmological constant leads to the occurence of
a cosmological horizon at $r=r_c$ (with $r_h < r_c < \infty$) where $f(r_c) = b(r_c)=0$.
This creates a difficulty since $r_c$ constitutes an apparent singular point of the equations.
In order to overcome this difficulty, we have solved the equations in two steps. In the
first step, we supplemented the system by the trivial equation $d \ell^2 /dx = 0$ and impose
the conditions of two regular  horizon at $r = r_h$ and $r=r_c$, fixing $r_c,r_h$ by hand
and solve the equations for $r \in [r_h,r_c]$.
The appropriate set of twelfe boundary conditions at the two horizons then read
\beq
        f = 0 \ , \ b = 0 \ , \ b' = 1 \ , \ w = w_h \ , V = 0 \ , \ a' = a'_h \ , \Gamma_1 = 0 \ , \ \Gamma_2 = 0
\  \ {\rm for} \ \ r = r_h
\eeq
\beq
        f = 0 \ , \ b = 0 \ ,   \Gamma_1 = 0 \ , \ \Gamma_2 = 0
\  \ {\rm for} \ \ r = r_c
\eeq
The functions $\Gamma_{1,2}$ are defined above;
here we take advantage of the arbitrary scale of the field $b(r)$ to impose $b'(r_h)=1$ and of the
arbitrary additive constant of the electric potential to assume $V(r_h)=0$. The function $b(r)$ will be renormalized
appropiately after the second step in order to set (\ref{asym}). 
The values $w_h$ and $a'_h$ are, a priori, arbitrary
and control the total angular momentum and the (electric and magnetic) charges of the black hole.

The numerical value of $\ell^2$ is determined by the first step, together with the values of
all the fields at $r=r_c$. Use of these fields and of the value of $\ell^2$ can then
be used as a suitable set of Cauchy data to solve the equations for $r \in [r_c,\infty]$. 
The disatvantage of the method is that we cannot perform a systematic analysis of the
solution for a fixed value of the cosmological constant. Fortunately, the numerical value
of $\ell^2$ depends only a little on $w(r_h)$ and $a(r_h)$ once $r_h,r_c$ are fixed. 

In the following, we present the results corresponding to $r_h=1,r_c=3$, this case corresponds
to a large value of the coupling constant but allows one to analyze the effect of $\Lambda$ on the
solution. We hope that the results for this case capture the main features of the solutions for
generic values of $\Lambda$ 
The profiles of the solution corresponding to  $w_h = 0.6, a'_h=0.5$
are presented in Fig. 1. It corresponds to  $1/\ell^2 = 0.0945$ and $q = -0.08538$.
The smoothness of the profile at $r=r_c$ can be appreciated on the plot. It is also
worth to point out that the numerical solutions approach the asymptotic behaviour
(\ref{asym}) although all boundary conditions are imposed at $r=r_c$ for the second step
of our construction.
We supplemented Fig. 1 with  the plot of the $tt$ component of the metric $g_{tt} = b - h w^2$,
 showind that there is a small region around the event horizon where $g_{tt}$ is negative,
defining an ergoradius $r_e$ where $g_{tt}(r_e)=0$ (on the figure, $r_e \approx 1.17$).
%%\newpage
\begin{figure}[!htb]
\centering
\leavevmode\epsfxsize=12.0cm
\epsfbox{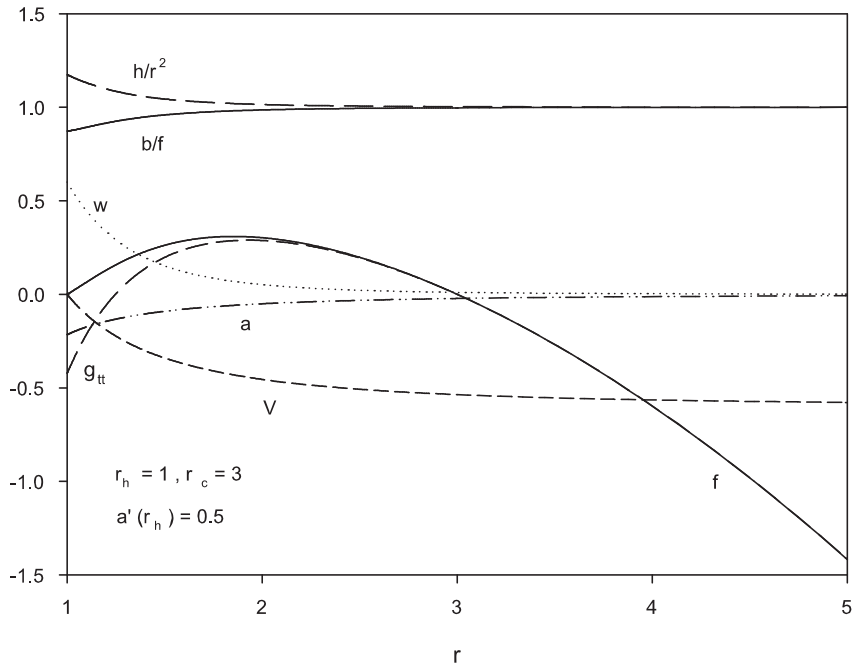}\\
\caption{\label{fig1} 
The profiles of a generic solution correponding to $r_h=1,r_c=3$ , $w_h = 0.6, a'_h=0.5$}
\end{figure}
%%\newpage
%%\vskip 3 cm
%%%%\newpage
\par
We manage to construct several branches of solutions for different values of $a'_h$ and increasing 
the parameter $w_h$.
For fixed $a'_h$, the solutions exist only for sufficiently large values of $w_h$, this is illustrated
on Fig.2 where we plot the asymptotic charges $M,J,Q,\mu$ as functions of $w_h$ for two values of
$a'_h$; for $a'_h = 0.1$ (resp. $a'_h = 0.5$) the branch exist for $w_h > 0.1 $ (resp. $w_h > 0.465$).
We strongly suspect that another branch of solutions, with a larger mass exist, terminating at the
same value of $w_h$ but we cannot construct it at this moment.  The numerical integration
between the  two horizons and the related constraints make the construction of an eventual second
branch very tricky.  We nevertheless observe that the asymptotic mass is positive for all values
of the parameters that we have explored. The figure further suggests that, if a second branch exists,
its asymptotic mass will be larger than the one of the first branch that we constructed. The magnetic moment
is rather independant on the parameter $w_h$. Also, when we solved the equation for larger values
of $r_c$, corresponding to smaller values of $\Lambda$, we observe that the asymptotic quantities
plotted on Fig. 2 depend weakly on $r_c$.
\\
\begin{figure}[!htb]
\centering
\leavevmode\epsfxsize=12.0cm
\epsfbox{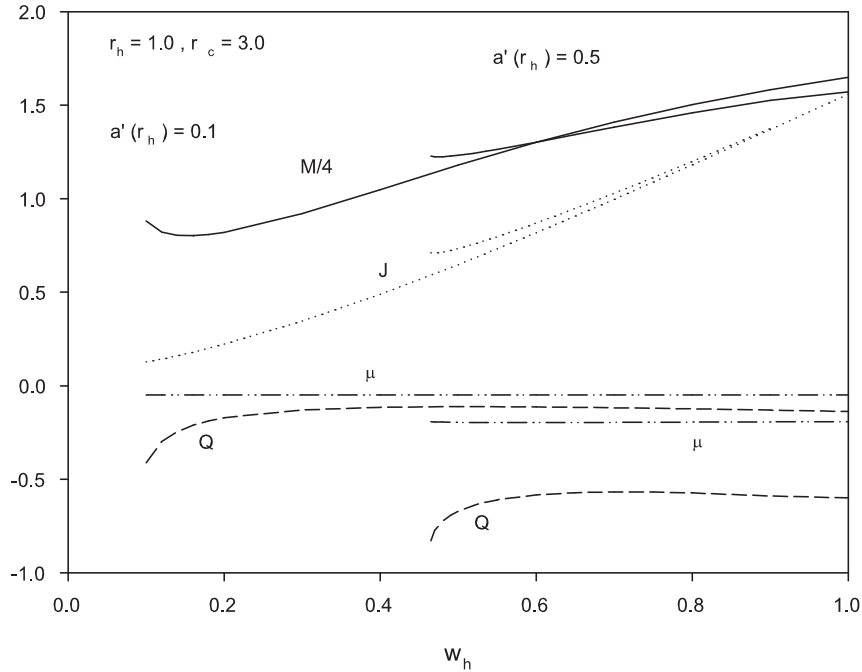}\\
\caption{\label{fig1} 
The asymptotic charges $M,J,Q, \mu$ are presented as functions of $w_h$ for $a'_h=0.1$ and $a'_h=0.5$}
\end{figure}
%%\newpage
%%\vskip 3 cm
The numerical values of the different fields at the horizon $r_h$ are presented on Figs. 3 and 4 
for the case $a'_h=0.1$ and $a'_h=0.5$ respectively.
The value $a_h$ is negative and depends weakly on $w_h$, e.g. we find $a_h \sim -0.049$
(resp. $a_h \sim -0.215$) for $a'_h=0.1$ (resp.  $a'_h=0.5$);  correspondingly we find
$\ell^2 \sim 10.3$ and $\ell^2 \sim  10.6$ ($\ell^2$ is indicated on Fig. 4 only).

\begin{figure}[!htb]
\centering
\leavevmode\epsfxsize=12.0cm
\epsfbox{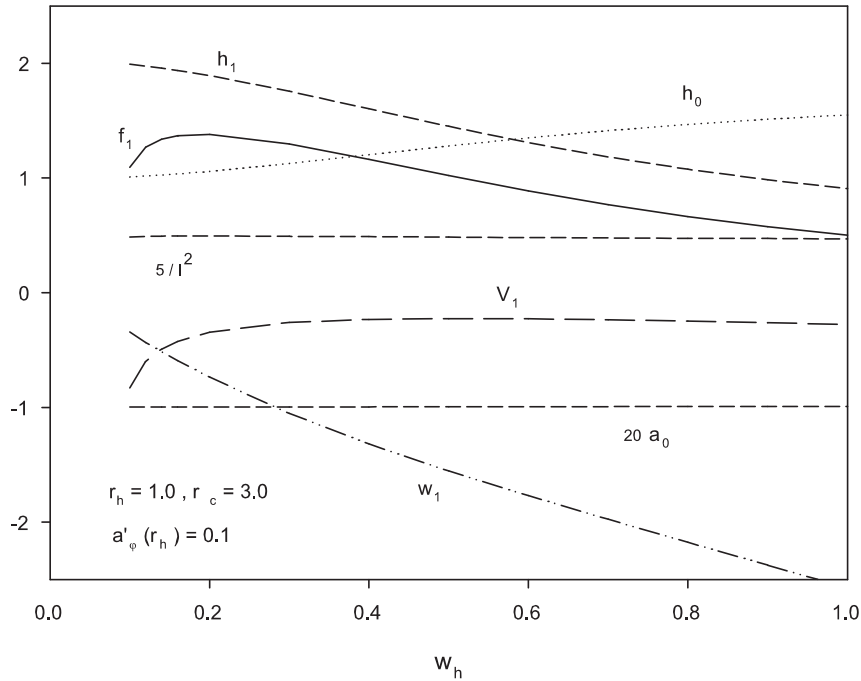}\\
\caption{\label{fig1} 
Some numerical parameters characterizing the solution at the horizon  as functions of $w_h$ for $a'_h=0.1$}
\end{figure}
%%%%\newpage
%%\vskip 3 cm
\begin{figure}[!htb]
\centering
\leavevmode\epsfxsize=12.0cm
\epsfbox{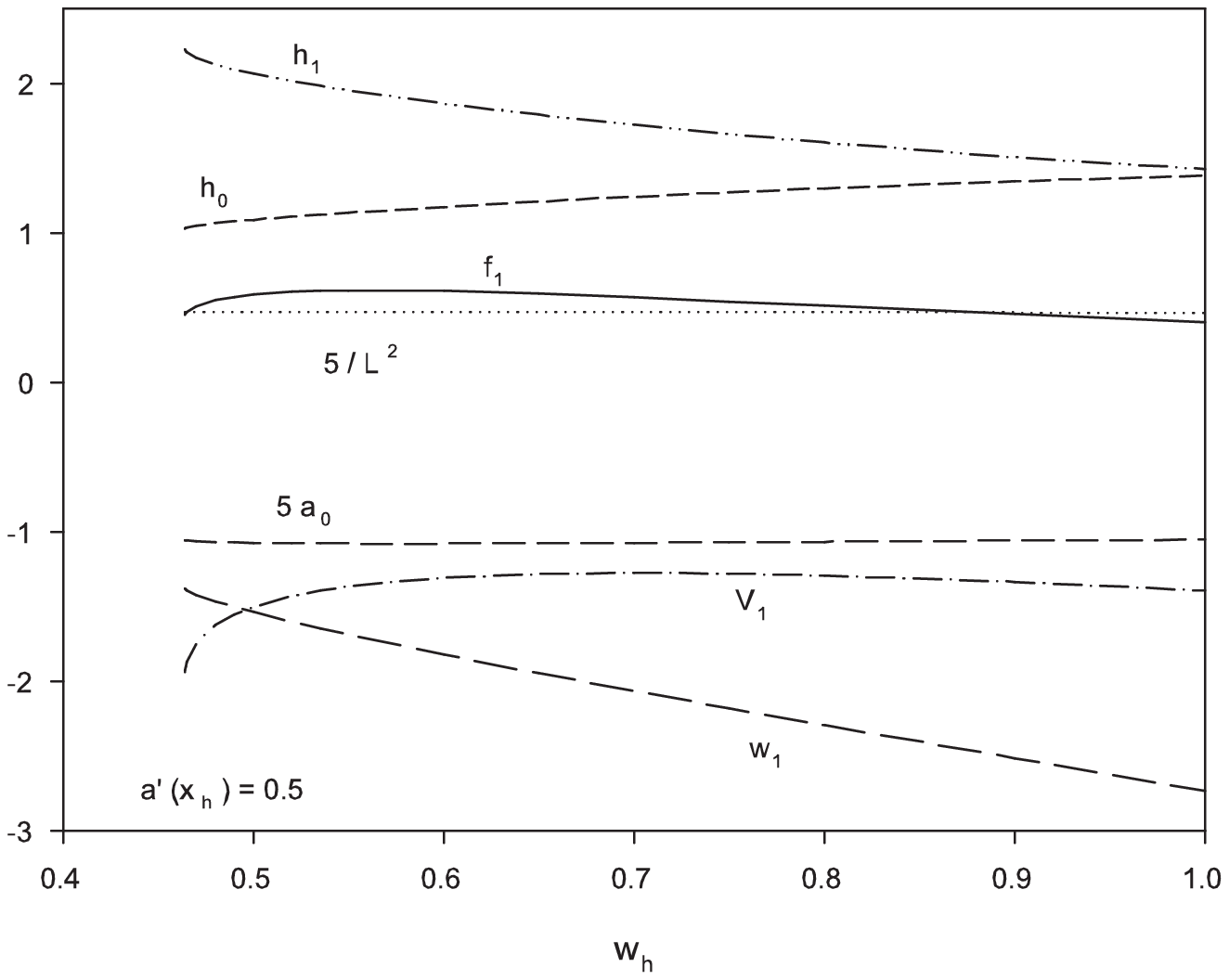}\\
\caption{\label{fig1} 
Some numerical parameters characterizing the solution at the horizon  as functions of $w_h$ for $a'_h=0.5$}
\end{figure}
\begin{figure}[!htb]
\centering
\leavevmode\epsfxsize=12.0cm
\epsfbox{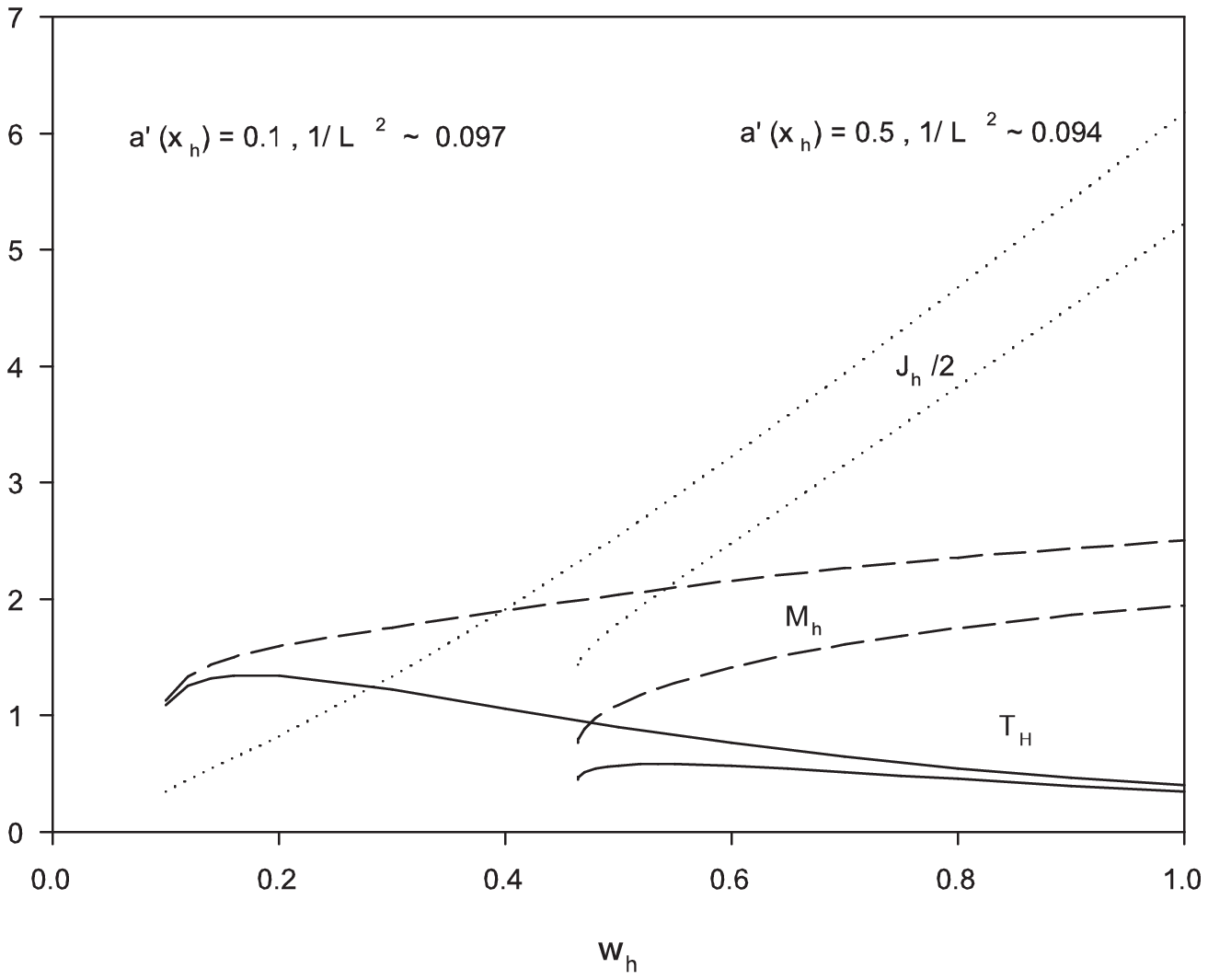}\\
\caption{\label{figxh_phy} 
The Hawking temperature at the event horizon $T_H$, the horizon mass $M_h$ and the horizon angular momentum
$J_h$  are represented respectively in solid, dashed and dotted lines as functions of $w_h$ for $a'_h=0.1$ and $a'_h=0.5$}
\end{figure}
\begin{figure}[!htb]
\centering
\leavevmode\epsfxsize=12.0cm
\epsfbox{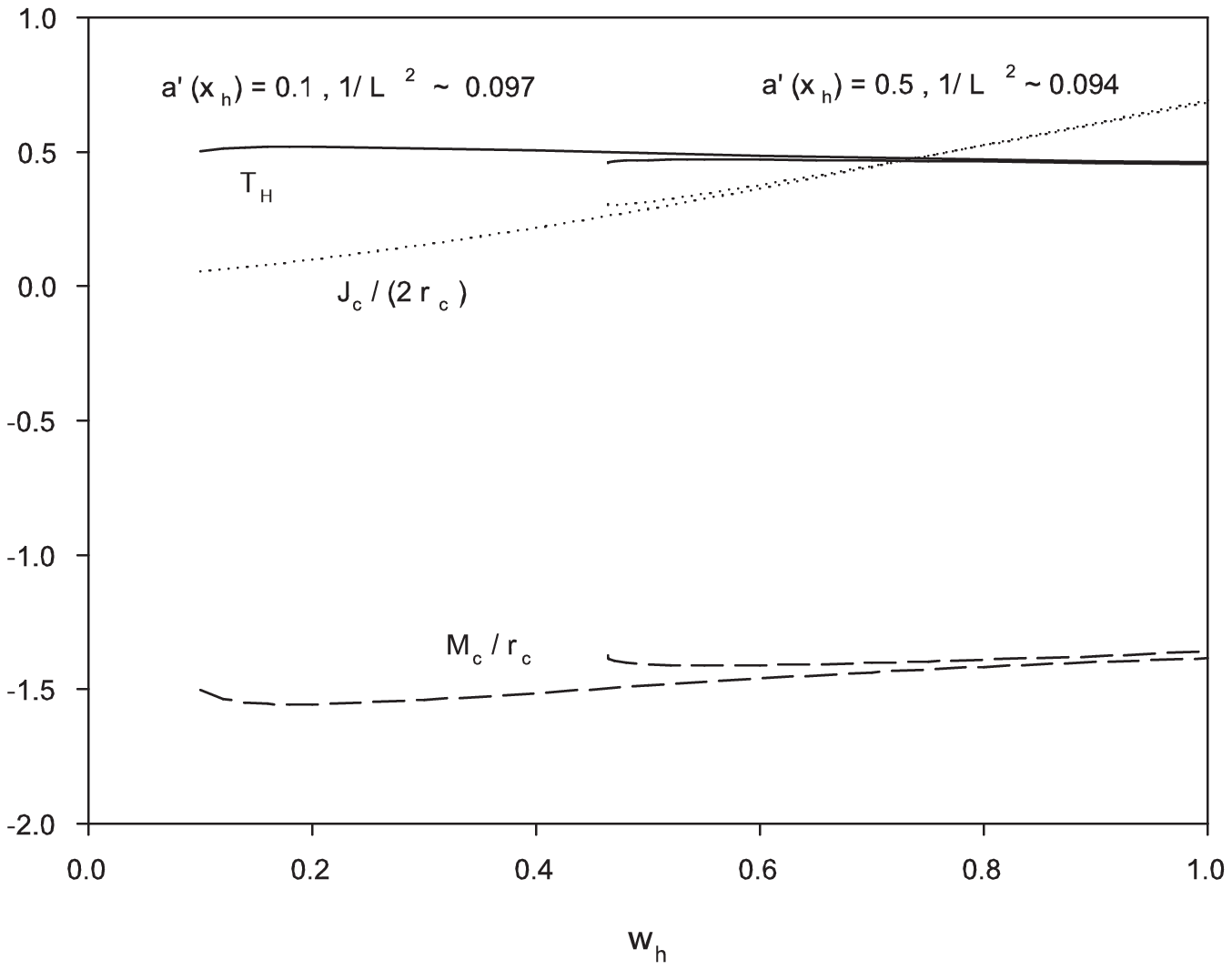}\\
\caption{\label{figxc_phy} 
The Hawking temperature at the cosmological horizon $T_H$, the horizon mass $M_c$ and the horizon angular momentum
$J_c$  are represented respectively in solid, dashed and dotted lines as functions of $w_h$ for $a'_h=0.1$ and $a'_h=0.5$}
\end{figure}

%%\newpage
%%\vskip 3 cm
%%%%%%%%%%%%%%%%%%%%%%%%%%%%%%%%%%%%%%%%%%%%%%%%%%%%%%%%%%%%%%%%%%%%%%%%%%%%%%%%%%%%%%%%%%%%%%%%%%%%%%%%
The physical parameters characterizing the solutions at the horizon can then be computed
from the numerical data. In particular, the natural normalisation of the function $b$ (i.e.
such that $b(r) \to - r^2/\ell^2$ for $r\to \infty$) which renders the space-time asymptoticall DeSitter  
is determined from the solution in the asymptotic region. On Fig. \ref{figxh_phy}, we have
superposed the Hawking Temperature $T_H$, the mass and angular momentum at the horizon
as functions of the parameter $w_h$ for the solutions obtained with $a'(x_h)=0.1$ (corresponding to
$1/\ell^2 = 0.097)$ and existing for $w_h > 0.1$ and  $a'(x_h)=0.5$ (corresponding to
$1/\ell^2 = 0.094)$ and existing for $h_h > 0.462)$. 
We see in particular that the Komar mass is positive at the even horizon.
For the sake of completeness, the same quantities relative to the cosmological
horizon $r_c$ are presented on Fig. \ref{figxc_phy}.  These physical parameters depend only weakly 
of $a'(r_h)$ when the parameter $w_h$ becomes sufficiently large. We note that the mass $M_c$ is negative
due to the fact that $b'(r_c)$ is negative and the rotation energy, although positive, is not
enough to make the combination in \ref{komar} positive at $r=r_c$. 
%%%%%%%%%%%%%%%%%%%%%%%%%%%%%%%%%%%%%%%%%%%%%%%%%%%%%%%%%%%%%%%%%%%%%%%%%%%%%%%%%%%%%%%%%%%%%%%%%%%%%%%%
%%%\vskip 3 cm
%%\newpage
%%%%\vskip 3 cm
%%%\vskip 3 cm
%%\newpage
%%%%\vskip 3 cm

%%%%%%%%%%%%%%%%%%%%%%%%%%%%%%%%%%%%%%%%%%%%%%%%%%%%%%%%%%%%%%%%%%%%%%%%%%%%%%%%%%%%%%%%%%%%%%%%%%%%%%%%
%%%%%%%%%%%%%%%%%%%%%%%%%%%%%%%%%%%%%%%%%%%%%%%%%%%%%%%%%%%%%%%%%%%%%%%%%%%%%%%%%%%%%%%%%%%%%%%%%%%%%%%%%
%%%%%%%%%%%%%%%%%%%%%%%%%%%%%%%%%%%%%%%%%%%%%%%%%%%%%%%%%%%%%%%%%%%%%%%%%%%%%%%%%%%%%%%%%%%%%%%%%%%%%%%%

\section{Black Strings}
Black strings solutions can also be constructed in the framework of the model (\ref{action-grav})
but we will limit our investigations to pure gravity in this case. One of the spacelike dimensions
of space-time, say $z \equiv x_{d-1}$ plays a special role in a sense that the metric 
is the warped product of a $d-1$-dimensional black hole metric with the extra dimension. The corresponding
horizon has the topology of $S_{d-3} \times S_1$ \cite{bs}.
In the case of non-rotating, uniform black strings, the fields do not
depend on the coordinate $z$ and the metric takes the form  
\begin{eqnarray}
\label{metricds} 
ds^2=a(r)dz^2+ \frac{dr^2}{f(r)}+r^2  d\Omega^2_{d-3} -b(r)dt^2
\end{eqnarray}
where $d\Omega^2_{d-3}$ denotes the metric on sphere $S^{d-3}$.
Solutions of the corresponding Einstein equations are constructed in \cite{mrs} for $\Lambda < 0$ 
and in \cite{bd} for $\Lambda > 0$. For $d$ even, the part of the metric related to the $d-3$-dimensional
sphere can be deformed according to the lines of Eq. (\ref{metric}) and rotating black strings can be
constructed.  The equations can be found in  Eqs.(4.2) 
of the recent \cite{brs} by changing $\ell^2 \to - \ell^2$ but we will write them here for completenes.
%%%%%%%%%%%%%%%%%%%%%%%%%%%%%%%%%%%%%%%%%%%%%%%%%%%%%%%%%%%%%%%%%%%%
A convenient metric gauge choice in the numerical procedure is $h(r)=r^2$. 
In this gauge, the field equations read~: 
\begin{eqnarray}
\label{eq1} 
\nonumber
\frac{a'}{a}=-\bigg[2\ell^2 
fg\left(rgb'+b(2g+(d-4)rg')\right)\bigg]^{-1} 
\bigg [ b\bigg(4(d-1)(d-2)rg^2
 -4(d-4)\ell^2g\big((d-2)r-fg'\big) \\
+(d-4)r\ell^2(4r^2+(d-5)fg'^2)\bigg)
+2 \ell^2 fg((d-4)rb'g'+g(2b'+r^3w'^2))\bigg], \nonumber
\end{eqnarray}
\begin{eqnarray}
\nonumber
f'=\frac{1}{d-2}\bigg(
\frac{(d-4)(2d-3)r^3}{g^2}
-\frac{(d-4)( d-2)r }{g }
-\frac{(d-1)(d-2)r }{\ell^2} \nonumber \\
+(\frac{ a'}{a}+\frac{b'}{b})((d-4)\frac{rg'}{2g}-d+3)f
\label{eq2} 
+\frac{ rfa'b'}{2ab}
-\frac{(d-3)(d-4)fg'}{g}   \nonumber \\
+\frac{(5-2d)r^3f}{2b}w'^2
+\frac{(d-5)(d-4)rf}{4g^2}g'^2 
 \bigg),
\end{eqnarray}
\begin{eqnarray}
\label{eq3}
\nonumber
b''=\frac{1}{d-2}\bigg(
\frac{(d-4)(d-5)b}{4g^2}g'^2
+\frac{(2d-3)r^2}{2}w'^2
-\frac{(d-3)(d-4) }{2g}b'g'
+\frac{(d-4) b}{2ag}a'g'
\\
-\frac{(d-2) }{2f}b'f'
+\frac{(d-2)b'^2}{2b} 
 \nonumber
-\frac{(d-3)}{2a }a'b'
+\frac{ba'}{ra} 
-\frac{(d-3)b' }{r} 
\nonumber \\
+\frac{(d-4) bg'}{rg} 
+\frac{(d-4)r^2b}{fg^2} 
-\frac{(d-2)(d-4)b}{fg} 
-\frac{(d-1)(d-2)b}{\ell^2f} 
\bigg)
\end{eqnarray}
\begin{eqnarray}
\nonumber
 g''=\frac{1}{d-2}\bigg(
\frac{r^2g}{2b}w'^2-\frac{d^2-7d+4}{4g}g'^2
-(d-2)\frac{f'g'}{2f}
+(\frac{a'}{a}+\frac{b'}{b})(\frac{g}{r}-g')
+\frac{ga'b'}{2ab}
-\frac{2g'}{r}
\\
\nonumber
+\frac{(4-3d)r^2}{fg}
+\frac{d(d-2)}{f}
-\frac{(d-1)(d-2)g}{\ell^2 f}
\bigg),
\end{eqnarray}
\begin{eqnarray}
\label{eqw}
\nonumber
 (g^{\frac{d-4}{2}} \sqrt{\frac{afh^3}{ b}}w')'=0~.
\end{eqnarray}
The last equation in the relations above implies the existence of the 
first integral
\begin{eqnarray}
\label{fiw}
w'=\alpha g^{-\frac{d-4}{2}} \sqrt{ \frac{b}{afh^3}},
\end{eqnarray}
where $\alpha$ is a constant controling the total angular momentum $J$ of 
the solutions. For the numerical analysis, we fix the arbitrary rescaling of the radial
variable by demanding $\ell^2 = 2000$.

For later use, we give here the expression of the Kretschmann invariant  with the 
metric (\ref{metricds}). The expression of this quantity turn out to be particularly simple~:
\be
\label{kret}
        K = \frac{d-3}{r^4} (r^2 (f')^2 + 2 (d-4) (f-1)^2)
\ee
%%%%%%%%%%%%%%%%%%%%%%%%%%%%%%%%%%%%%%%%%%%%%%%%%%%%%%%%%%%%%%%%%%%%%%%%%%%%
\\
\subsection{Boundary Conditions and asymptotic behaviour}
\par In order to construct black string solutions with  the equations above, we have to impose the appropriate
boundary conditions at the horizon $r=r_h$ which we require to be regular. For this purpose, we set
\be
\label{localbs}
f(r_h) = 0 \ , \  b(r_h) = 0 \ , \ a(r_h) = 1 \ , \ b'(r_h) = 1 \ , \ w(r_h) = w_h 
\ee
plus another condition ensuring that the equation for $g$ is regular at the horizon that we do not write here
(it is Eq.(4.5) of \cite{brs}).
 The first condition  above is necessary to produce a "`black"' object, the second one is necessary for regularity of the equation for $b$. The third and fourth conditions fix the arbitrary scales of the functions $a$ and $b$. 
 Finally the last condition involves an arbitrary parameter $w_h$ which control the angular velocity
 of the black string at the horizon.
 The Horizon mass and angular velocity of the solution can, again, be determined with suitable Komar integrals, leading to 
 \be
  M_h = \sqrt{\frac{f}{b h g^2}} (b'  - h w w') \ \ , \ \     J = 2 \sqrt{\frac{a f g^2 h^3}{b}} w'\vert_{r=r_h}
 \ee
  \\
The complete specification of the boundary values needs three extra conditions which have to be
looked for in the asymptotic region.
%%%\subsection{Asymptotic expansion}
Examining the asymptotic behaviour compatible with the classical equations reveals 
at least two possibilities. The first one corresponds to the metric of a  De Sitter space-time
\beq
\label{desitteras}
a(r) = - \frac{r^2}{\ell^2} \ \ , \ \   
b(r) = - \frac{r^2}{\ell^2} \ \ , \ \
f(r) = - \frac{r^2}{\ell^2} \ \ , \ \
g(r) = r^2 \ \ , \ \ 
w(r) = J (\frac{l}{r})^{d-2}
\eeq 
more details about the corresponding asymptotic expansion of this solution are given  in \cite{brs}.
Correspondingly, the Kretschmann invariant approached a constant asymptotically. 
However (\ref{desitteras}) is not the only possibility, there exist
a second one where the fields decay  power likely according to 
\beq
\label{kasneras}
a(r) =  A (\frac{r^2}{\ell^2})^{\alpha/2} \ \ , \ \   
b(r) =  B (\frac{r^2}{\ell^2})^{\beta/2} \ \ , \ \
f(r) =  F (\frac{r^2}{\ell^2})^{\phi/2} \ \ , \ \
g(r) = r^2 \ \ , \ \ 
w(r) = \Omega \ \ell^4 (\frac{\ell^2}{r^2})^{\omega/2}
\eeq  
with 
\beq
\label{exposant}
       \alpha =  \beta = -2(d-3) - \sqrt{2(d-2)(d-3)} \ , \ \  
       \phi = 2(d-2) + 2 \sqrt{2(d-2)(d-3)} \ \ , \ \ 
       \omega = 2 + \frac{\phi}{4}
\eeq
This leads to a singularity of the Kretschmann invariant in the limit $r \to \infty$ for $d>3$. 
More details about the asymptotic expansion of (\ref{kasneras}) are presented in the next section. 

%%%%%%%%%%%%%{\bf PARTIE TERENCE}

\subsection{Non rotating solution}
\par In the case $w=0$, the equation for $w$ is trivial and the equation for $g$ is satisfied by $g=r^2$. 
We will show that the remaining equations can be written in a decoupled form. We consider 
both $\Lambda>0$ and $\Lambda<0$ ; to this use, we define $\epsilon = -\Lambda/|\Lambda|$.\\

\par First, let us define 
\be
\label{eqter1}
A(r) = \frac{ra'}{a}\ \ B(r) = \frac{rb'}{b}
\ee
in terms of which the equations for $f,a,b$ rewrite~:
%%%%\par Then, equations \eqref{eq2} become
\be
\label{eqter2}
rf' = 2(d-4)\left(k-f\right) - 2(d-1)\frac{r^2}{\epsilon l^2} - f(r)\left(A+B\right)
\ee

\be
\label{eqter3}
rB'f = (d-1)(2-B) \frac{r^2}{\epsilon l^2} - (d-4)kB = 0
\ee

\be
\left(AB + 2(d-3)(A+B) - 2(d-3)(d-4)\right)f 
= 2k(d-3)(d-4) - 2(d-1)(d-2)\frac{r^2}{\epsilon l^2}
\ee

\par  Solving the third equation for $A = A(B,f)$, substituting $A(B,f)$ in the first equation and solving for $B=B(f,f')$ and finaly evaluating the second equation with $A=A(B(f,f'),f),\ B=B(f,f')$ leads to the following decoupled equations~:

\bea
r^2ff'' &=&  - rff' + \left(rf'\right)^2 + 2\left( (d-1) - (d-4)\frac{r^2}{\epsilon l^2}\right)^2 \\
&-& 2(d-4)\left( (d-1) - (d-4)\frac{r^2}{\epsilon l^2}\right)f - 3\left( k(d-1) - (d-4)\frac{r^2}{\epsilon l^2}\right)rf'\nonumber
\label{eqf}
\eea

\bea
B_\pm &=& -\tilde B(r,d,l) \\
      &\pm& \sqrt{ \tilde B(r,d,l)^2 - 2(d-3)(d-4)\left(1-\frac{1}{f} \right) - 2(d-3)\frac{rf'}{f} - 2(d-1)(d-4)\frac{r^2}{\epsilon l^2}\frac{1}{f}}\nonumber
\label{eqB}
\eea

where $\tilde B(r,d,l) \equiv \left( (d-4) + \frac{1}{2}\frac{rf'}{f} + (d-4)\frac{1}{f} - (d-1)\frac{1}{f}\frac{r^2}{\epsilon l^2}\right)$
\be
A = \frac{-2(d-1)(d-2)\frac{r^2}{\epsilon l^2} + 2(d-3)(d-4)f + 2(d-3)fB}{2(d-3)f + fB}
\label{eqA}
\ee
\par Note that the decoupling is still valid in the more general case considered in \cite{mrs} where
topological black holes are investigated as well (i.e. space-times  where the spherical part $d\omega^2_{d-3}$
 in (\ref{metricds}) is replaced by the metric of an hyperbolic or flat manifold with the same dimensions)

%%%%%%%%%%%%%%%%%%%%%%%%%%%%%%%%%%%%%%%%%%%%%%%%%%%%
% ASYMPTOTICS
%%%%%%%%%%%%%%%%%%%%%%%%%%%%%%%%%%%%%%%%%%%%%%%%%%%%

\subsection{Asymptotics}

We have not been able to solve \eqref{eqf} explicitely, but we can study the asymptotic behavour of $f$ with this equation. In fact, at first order, it appears that
\be
f(r)\approx F_0r^\phi
\ee
for a constant $F_0$ and for any $\phi \in \{2\}\cup]4,\rightarrow$, so we cannot conclude on the asymptotic behavour of $f$ at this point. In fact, it is not possible to fix analytically $\phi$, but it is possible to give a description of the asymptotic behavour of the metric functions in terms of only one parameter.\\

\par We have to consider two cases : $\phi = 2$ and $\phi>4$. In the case $\epsilon = +1$, where the cosmological constant is negative, asymptotics obey $\phi=2$ \cite{mrs}. In the case $\epsilon=-1$, there are strong evidence that $\phi=2$ leads to singular solutions \cite{bd}. Since this paper focus on the case $\epsilon=-1$, we will assume $\phi>4$, which is equivalent to assume that the argument in five dimensions generalises to $d$ dimensions.\\

\par It turns out that the asymptotic behavour of the metric functions is given by a one parameter family of functions:
\be
f\rightarrow F_0r^\phi\ , \ \phi = -\frac{\beta^2 + 2(d-4)\beta + 2(d-3)(d-4)}{\beta + 2(d-3)}
\ee
\be
A(r)\rightarrow \alpha(\beta) = -\frac{2(d-3)(d-4) + 2(d-3)\beta}{\beta + 2(d-3)}
\ee
\be
B(r) \rightarrow \beta
\ee

\par Since these behavour are compatible with the equations of motion for 
each value of $\beta$ such that $\phi > 4$, it is not possible to fix $\beta$ without more asomptions, but there are strong numerical evidences for $a(r) = b(r)$ at first order in the asymptotic region. %%%\rouge{INSERER FIGURE OU CITER PAPIER A 5 DIM} 
This means that the value of $\beta$ chosen by the system is a fixed point of $\alpha(\beta)$
are exactly the values (\ref{exposant}).
%%
%%\be
%%\beta = -2(d-3) - \sqrt{2(d-2)(d-3)}
%%\ee
%%with this value of the parameter, we have
%%\be
%%phi = 2(d-2) + 2\sqrt{2(d-2)(d-3)}
%%\ee
Note the particular relation between the exponents~: 
\be
\alpha + \beta + \phi + 2(d-4) = 0
\label{link}
\ee

We have constructed the higher order corrections for the function $f(r)$ and obtained
\be
f(r) = F_0r^\phi - \frac{d-1}{d-3+\beta} \frac{r^2}{l^2} + \frac{d-4}{d-4+\beta}
\label{secord}
\ee
with $\phi$ discussed previously. The reparametrisation of $F_0$ used in (\ref{kasneras}) 
figure out the peculiar dependance of the solution under rescaling of $\ell$ and $r$   \\

%%%%%%%%%%%%%%%%%%%%%%%%%%%%%%%%%%%%%%%%%%
\subsection{Energy of the solution}
The energy of the solution is given by the following quantity \cite{Gibbons:1976ue}
\be
E = -2M_P^{d-2}\int_{S_t^\infty} N\left( K_{d-2} - K_{0,d-2} \right)
\label{masshawk}
\ee
where $M_P$ is the Planck Mass, $N$ being the full spacetime lapse function.\\

\par The trace extrinsic curvature of the border at fixed time, $K_{(d-2)}$ is well defined, although
space-time is asymptotically singular~:
\be
K_{(d-2)} = \sqrt{f}\left(  \frac{(d-3) + \frac{\beta}{2}}{r}\right)
\ee
This quantity, once integrated according to \eqref{masshawk}, is not divergent.  That's the reason why we won't consider a reference background. Moreover, it is not clear wich background to use with such an asymptotic.\\

\par We can now compute the energy of the solutions which exhibits an 
asymptotic behavour as described in the previous section. By \eqref{link}, the energy is simply given by
\be
E =  M_P^{d-2}\mathcal A_{d-3} L \sqrt{ 2F_0(d-2)(d-3)}
\ee
where $L$ is the length of the extradimension in the $y$ direction and $\mathcal A_{d-3}$ is the area of the unit $d-3$ sphere. This energy is finite and depend only on $F_0$ wich depends essentially on the  dimension of space-time and of the cosmological constant, further results will be given in  Sect. 3.7.

%%%%%%%%%%%%%%%%%%%%%%%%%%%%%%%%%%%%%%%%%%%%%%%%%%%%%
%%%% Geodesic equations and Curvature invariants  %%%
%%%%%%%%%%%%%%%%%%%%%%%%%%%%%%%%%%%%%%%%%%%%%%%%%%%%%
\subsection{Geodesic Equations and Curvature Invariant}
We computed the Kreshman curvature invariant (\ref{kret}):
%%\be
%%K = (d-3)\frac{2(d-4)-4(d-4)f +2(d-4)f^2 + r^2f'^2}{r^4}
%%\ee
and it turns out that the spacetime is asymptotically singular, with the asymptotic behavour of previous section.\\
 Moreover, the geodesic equation at fixed $\theta_i$ (the angular sector of the metric), 
\be
\dot r^2 = \frac{\mathcal E^2}{b} - \frac{\mathcal Z^2}{a} - m^2
\ee
where $\mathcal E, \mathcal Z$ are conserved quantities along the geodesic, implies that this singularity can be reached in a finite proper time for an observer of mass $m$ from outside the horizon. A similar result holds  for null motion.
%%%%%%%%%%%%%%%%%%%%%%%%%%%%%%%%%%%%%%%%%%%%%%%%%%%%%%%%%%%%%%%%%%%%%%%%%%%%%%%%%%%%%%%%%%%%%%%%%%%%%%%%%%%%%%%%%%%%%%%%%ùùù

\subsection{Numerical Results}

\par We solved numerically equations \eqref{eqter1}, \eqref{eqter2} and \eqref{eqter3} 
with the boundary conditions for many values of the horizon $r_h$.
It turns out that in the regular case, the asumption $a=b$ holds everywhere, independently of the number of dimensions. This can be understood since $a$ and $b$ both play a "spectator role" in the metric. Nothing explicitely depends on $z$, neither on $t$ in the metric, so in a sense, $a$ and $b$ play the same role, at least in the regular case, since the initial values for $a$ and $b$ are the same
On fig.6, we present the evolution of the ratio $rf'/f$ for non rotating black string
solutions and for several values of $d$ and $r_h=0.5$ . The
figure clearly demonstrates that the power law configuration (\ref{kasneras}) is approached.
The exact values of the exponents coincide with our numerical values within the numerical
accuracy required for the numerics, i.e. typically $10^{-8}$.

%%\rouge{Demander a Radu qu'il vérifie l'explication puisqu'on cite son nom} \footnote{The author thanks E. Radu for this %%remark}.\\
%%%\vskip 6 cm
\begin{figure}[!htb]
\centering
\leavevmode\epsfxsize=12.0cm
\epsfbox{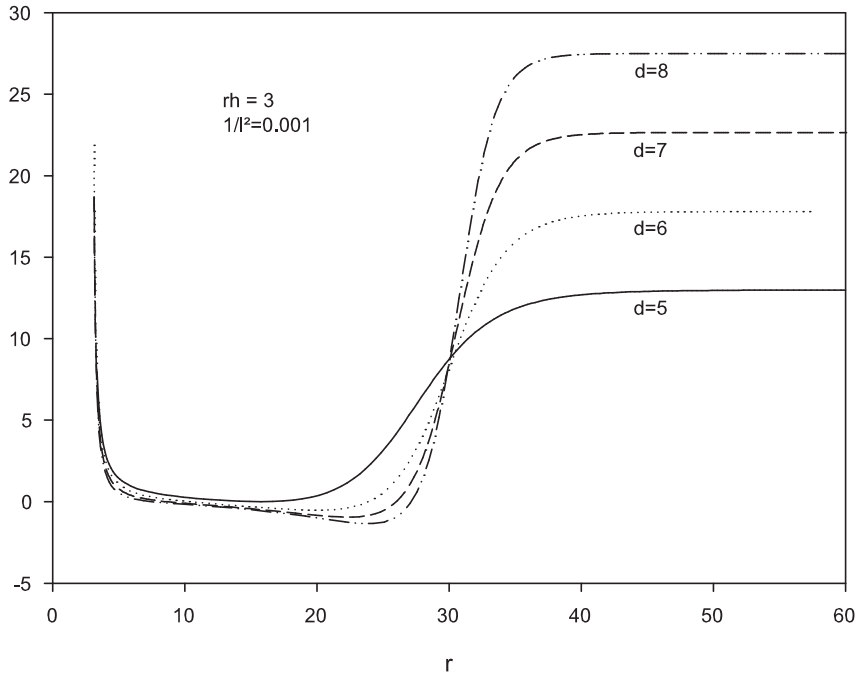}\\
\caption{\label{fig6} 
The ratio $xf'/f$ is given for several values of $d$}
\end{figure}
%%%\newpage

\par However, nothing garanties that this asumption will still hold true in the black string case, because the boundary conditions break this "symetry". We checked numerically the validity of the asumption $a=b$ in the asymptotic region and it turned out that it is still verified.
%%\rouge{FIGURE}. 
Moreover, we noticed that the ratio $a/b$ behaves like $1/r_h$ for the normalisation of $b$ we used. This factor can be absorbed in a rescalling of $t$. It is reasonable to think that a natural rescaling of $t$ is such that $a/b\rightarrow C$, with $C$ independant of $r_h$. This implies that $b'(r_h)=f'(r_h)$ wich reminds the 4-dimensionnal Schwarzchild (with or without cosmological constant).  \\

We also investigated the solution in the interior region, i.e. for $r \leq r_h$, to check whether there
exist a second horizon which would "`hide the asymptotic singularity"', but it turns out that this is
not the case. We have a naked singularities both, at $r=0$ and at $r=\infty$. We are tempted to interpret
this solution as an hypercylindre with an horizon at some equator by matching the origin with the point
at infinity by means of an appropriate system of coordinates, still to be found, but  that we expect to exist.

%%\par We also computed the coefficient $F_0$ for many values of $\Lambda$ and observed the following behavour, for a given %%number of dimension and horizon 
%%\rouge{Figure}:
%%\be
%%F_0 = \Lambda^{\frac{\phi}{2}}\tilde F_0(d,r_h)
%%\ee
%%This is simply due to an invariance of the equation under
%%\be
%%r\rightarrow r/l = \sqrt{\Lambda}r
%%\ee

%%%%%%%{\bf FIN PARTIE TERENCE}
%%%%%%%%%%%%%%%%%%%%%%%%%%%%%%%%%%%%%%%%%%%%%%%%%%%%%%%%%%%%%%%%%%%%%
\subsection{Rotating black strings} 
\begin{figure}[!htb]
\centering
\leavevmode\epsfxsize=12.0cm
\epsfbox{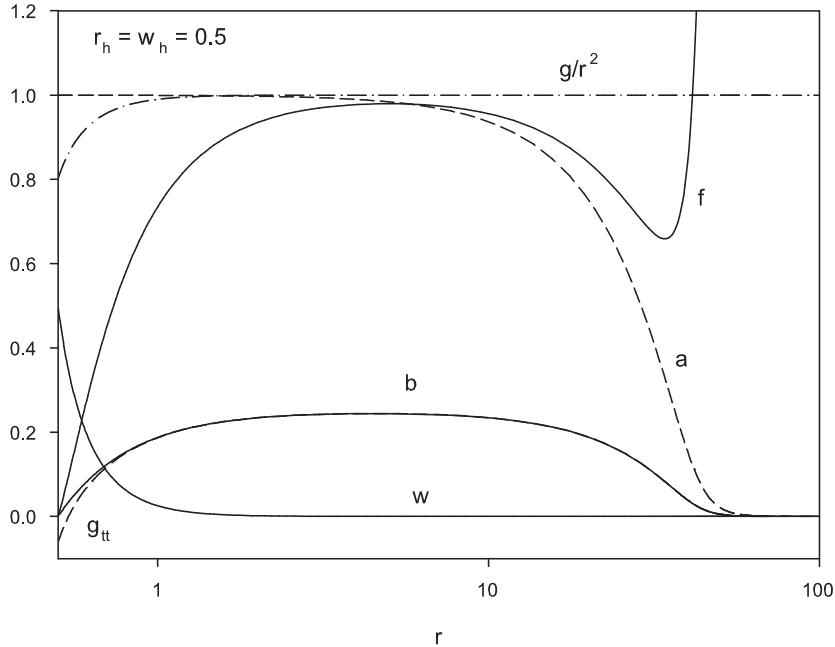}\\
\caption{\label{rota0} 
The profiles of a rotating black string for $r_h=w_h=0.5$}
\end{figure}

In absence of explicit solutions,
we have integrated numerically the equations for rotating black strings
for $r\in [r_h, \infty]$ for several values of $r_h$ and $w_h$ by trying to interpolate between
the local behaviour (\ref{localbs}) and one of the possible asymptotic behaviours
(\ref{desitteras}) or (\ref{kasneras}). 
Here we solved the equations for $d=6$ but the results obtained for non rotating black string for $d>6$
strongly suggest that rotating solutions exist also for higher dimensions.
As can be expected from the
non rotating case, our numerical results strongly suggest
that the rotating solutions behaving regularly at the event horizon naturally evolve into
the asymptotics determined by (\ref{kasneras}). The profiles of a rotating black string
corresponding to $r_h=0.5,w_h=0.5$ is presented of Fig. \ref{rota0}. The 
metric functions profile are shown on this figure.  We supplemented $g_{tt}= b - r^2 \omega^2$, showing
that there is a small ergoregion about the horizon where $g_{tt} < 0$ for $r_h \leq r \leq r_e$ where $r_e$
denotes the ergo-horizon (on the figure $r_e \approx 0.537$). For this solution we further found
$F\approx 1.125$, $A\approx 0.0741$, $B \approx 0.0186$ and $ \Omega \approx 0.0073$ for the parameters
 defined in (\ref{kasneras}).
%% \vskip 3 cm
\par
\begin{figure}[!htb]
\centering
\leavevmode\epsfxsize=12.0cm
\epsfbox{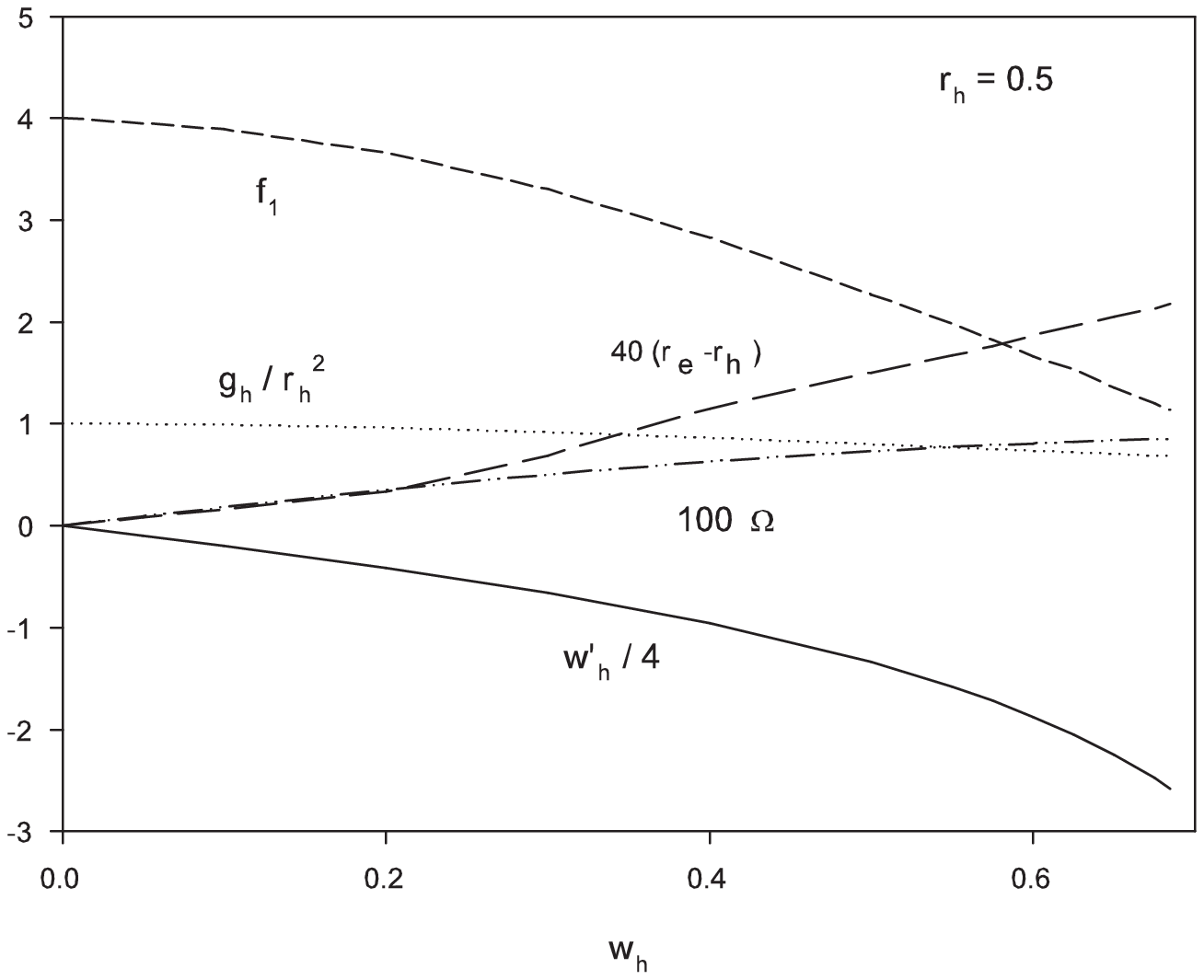}\\
\caption{\label{rota1} 
The values of $g,f',w'$ at the horizon, of the ergo radius and the parameter $\Omega$  are presented
as functions of $w_h$ for $r_h=0.5$}
\end{figure}

 We also studied the dependance of different parameters characterizing the solution 
as functions of $r_h$ and $w_h$.
One main feature is that the values of $F,A,B$ depend weakly of these parameters
(e.g. for $w_h=0.5, r_h = 0.8$ we find $ F\approx 1.125, A \approx  0.0738, B \approx  0.0260$).
The dependance of $\Omega$ on $w_h$ is more sensible. On fig. \ref{rota1} we have superposed
$\Omega$, $f_1$, $w'_h$ and the ergo-horizon $r_e$ as functions of $w_h$ for $r_h=0.5$.
The qualititive behaviour of these parameters remains similar for different values of $r_h$.
We noticed that, while increasing $w_h$, the function $w(r)$ becomes very peaked at the horizon,
with a large derivative $w'(x_h)$. This renders the numerical integration tricky in this region
but we have not detected a signal of a second branch of rotating black string.

%%%\newpage
%%\vskip 3 cm
%%%\newpage
%%\vskip 3 cm

Before finishing this section, we would like to mention that
an analytic solution of the Lagrangian under consideration is available in the case $d=4$
 \cite{linet}, see also  \cite{bbh} where this solution is interpreted 
in the context of gravitating cosmic strings. This solution was obtained
  with a different parametrisation of the metric, the correspondic space-time becomes
periodic in the radial variable; the Kretschmann scalar possesses 
singularities, so this solution is clearly not of the DeSitter type asymptotically.
%%%% 
%%%%%%%%%%%%%%%%%%%%%%%%%%%%%%%%%%%%%%%%%%%%%%%
\section{Summary}
In this paper, we have studied the Einstein-Maxwell equations in space-times of arbitrary dimensions $d$
and with a positive cosmological constant. By using appropriate ansatzes for the metric and the U(1)-fields,
the equations can be transformed into a system of ordinary differential equations 
for odd values of $d$ in the case of black holes and even $d$ for black strings.
Imposing a consistent set of boundary conditions, we solved these equation numerically,
constructed several  families of rotating black holes and black string solutions.   
For both cases, the conditions of an event horizon are imposed ar $r = r_h$. The main difference
between black holes and black strings resides in the way space-times reaches its asymptotic form.
In the case of black holes, space-time becomes asymptotically DeSitter, after crossing a regular
cosmological horizon. In the case of black strings, our numerical results strongly suggest that
the metric fields  evolve asymptotically according to some power of the radial variable $r$ with
well specified non integer exponants depending on $d$. The evaluation of the Kretschmann invariant
reveals that space-time become singular in the limit $r\to \infty$ for $d\geq 5$.
In order to confirm these numerical results by an analytic argument, 
we manage to  put the equations of \cite{Copsey:2006br,mrs} in a decoupled form which makes slighly easier
to construct the next to leading order of the asymptotic evolution of the fields. Althought we were not able to give an exact solutions to this decoupled form, we found interesting informations on the behavour of the solutions.   
We also showed that the energy of this kind of solution is finite.
\\
\newpage
{\bf Acknowledgments} Y.B. acknowledges B. Hartmann for early discussions on the topic of black strings. 
We gratefully acknowlegde E. Radu for numerous discussions and for constructive remarks about   
the manuscript.
\\
%%\newpage
%%%%%%%%%%%%%%%%%%%%%%%%%%%%%%%%%%%%%%%%%%%%%%%%%%%%%%%%%%%%%%%%%%%%%%%%%%%%%%

\end{document}